\documentclass{iau} 
\usepackage{graphicx}

\title[IAU Symp.\ 316.~~Mapping out the origins of compact stellar systems] 
{Mapping out the origins \\ of compact stellar systems}

\author[Aaron J.\ Romanowsky, Jean P.\ Brodie, and the SAGES]   
{Aaron J.\ Romanowsky,$^{1,2}$
 Jean P.\ Brodie,$^2$
 \and the SAGES$^3$}

\affiliation{$^1$Department of Physics \& Astronomy, San Jos\'e State University, \\ San Jose, CA
95192, USA \\ email: {\tt aaron.romanowsky@sjsu.edu} \\[\affilskip]
$^2$University of California Observatories, \\ 1156 High Street,
Santa Cruz,CA 95064, USA \\email: {\tt jbrodie@ucsc.edu}
 \\[\affilskip]
 $^3${\tt http://sages.ucolick.org}
}

\pubyear{2015}
\volume{316}  
\jname{Formation, evolution, and survival of massive star clusters}
\editors{C.\ Charbonnel \& A.\ Nota, eds.}
\begin{document}

\maketitle

\begin{abstract}
We present a suite of extragalactic explorations of
the origins and nature of globular clusters (GCs) and ultra-compact dwarfs (UCDs),
and the connections between them.
An example of GC metallicity bimodality is shown to reflect underlying, distinct metal-poor and metal-rich stellar halo populations.
Metallicity-matching methods are used to trace the birth sites and epochs of GCs in giant E/S0s,
pointing to clumpy disk galaxies at $z\sim3$ for the metal-rich GCs, and
to a combination of accreted and in-situ formation modes at $z\sim 5$--6 for the metal-poor GCs.
An increasingly diverse zoo of compact stellar systems is being discovered, including objects
that bridge the gaps between UCDs and faint fuzzies, and between UCDs and compact ellipticals.
Many of these have properties pointing to origins as the stripped nuclei of larger galaxies,
and a smoking-gun example is presented of an $\omega$~Cen-like star cluster embedded in a tidal stream.
\end{abstract}

\firstsection 
\section{Introduction}

Globular star clusters (GCs) are spectacular residents of our Galaxy's halo and beyond, yet there are
long standing mysteries as to how, when, and where they formed.
These mysteries expanded with the recent discoveries of ultra-compact dwarfs (UCDs), 
which began to bridge the classical gap between galaxies and star clusters
and raised questions of how to differentiate between these two classes of stellar system.
While studies of GCs in the Milky Way (MW) are focusing on remarkable clues about their origins
from multiple stellar populations, 
another important window is to examine the extragalactic populations of GCs and UCDs -- 
to provide additional context, statistics, and opportunities for witnessing formation in action.

Here we discuss briefly three areas of novel discovery and insight
emerging recently from extragalactic studies.
The first is new constraints on GC origins in a cosmological context  (\S\ref{sec:cosmo}), 
the second is more extensive mapping of the parameter space of compact stellar systems (\S\ref{sec:CSS}),
and the third concerns observations of tidally stripped nuclei (\S\ref{sec:inform}).

\section{Globular cluster bimodality and cosmological origins}\label{sec:cosmo}

GCs are thought to trace ancient phases of star and galaxy formation,
but many of the particulars are murky.
The MW provides important clues through the {\it bimodality} of its GC system,
which consists of a roughly spherical, non-rotating, metal-poor ``halo'' component, and a
flattened, rotating, metal-rich, ``bulge/thick-disk'' component.
This configuration suggests a early, chaotic assembly through infalling dwarf galaxies
(e.g., \cite[Searle \& Zinn 1978]{Searle78}), 
followed by intense star-forming activity in the main part of the Galaxy.
A similar pattern is seen in other galaxies, including the giant ellipticals and lenticulars (E/S0s),
which commonly show color bimodality (blue and red GCs; e.g., 
\cite[Zepf \& Ashman 1993]{Zepf93}; \cite[Larsen et al.\ 2001]{Larsen01}).
Despite some controversies about interpreting the colors,
bimodality in metallicity has now been confirmed through spectroscopic efforts with Keck/DEIMOS
as part of the SLUGGS survey\footnote{\tt http://sluggs.ucolick.org} 
(Brodie et al.\ 2012, 2014; Usher et al.\ 2012),
with supporting evidence from kinematics (Pota et al.\ 2013).

\begin{figure}[h]
\begin{center}
 \includegraphics[width=4.5in]{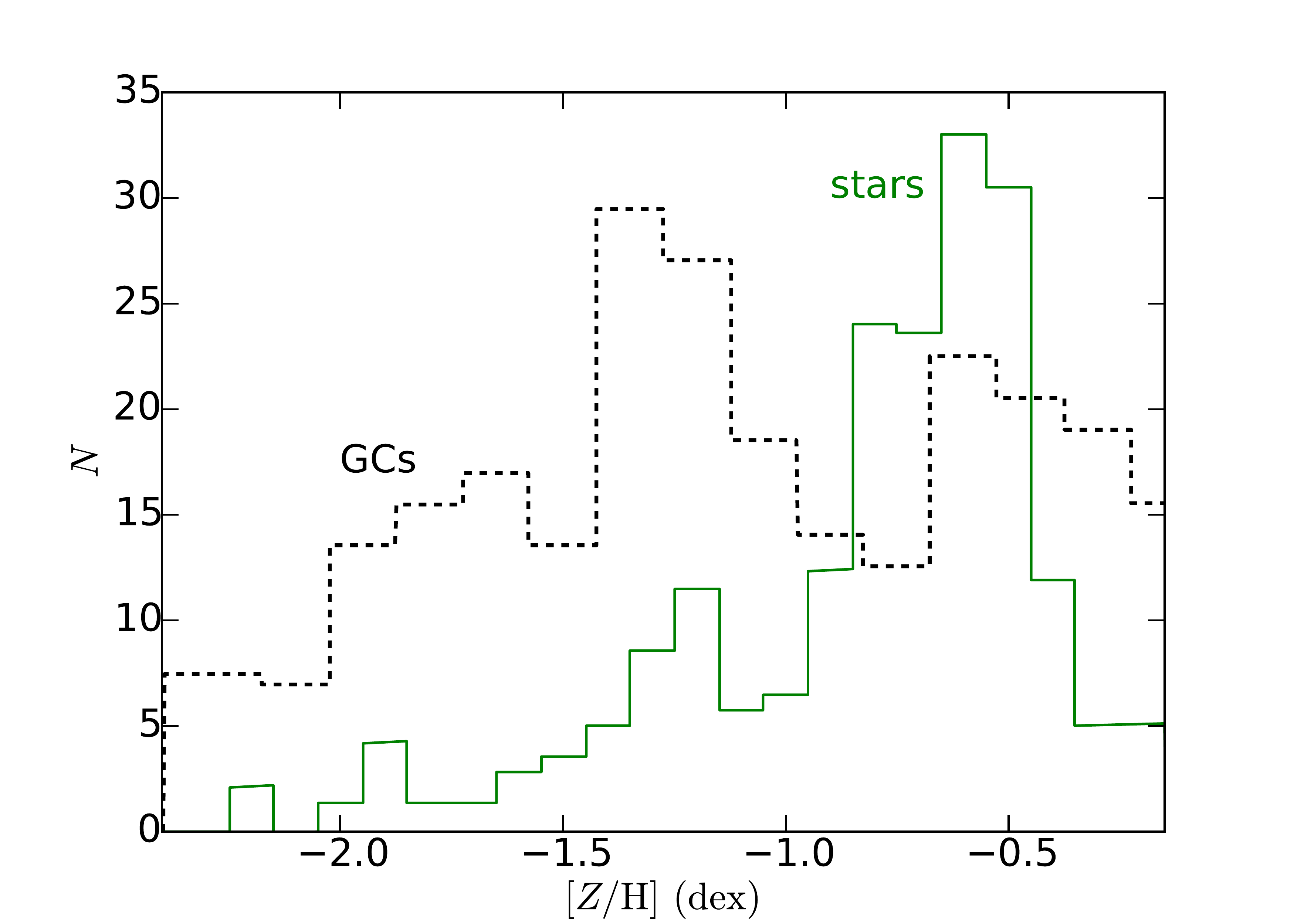} 
 \caption{Metallicity distributions of GCs (black dashed histogram) and halo stars (green solid histogram) around the S0 galaxy NGC~3115 
 (\cite[Peacock et al.\ 2015]{Peacock15}).
 The stars are from deep {\it HST} imaging of the RGB at projected galactocentric radii of $\sim$~50--60~kpc, and the GCs
 from Subaru/Suprime-Cam photometry at comparable radii (spectroscopic metallicities at smaller radii show similar results).
 The stellar histograms have been arbitrarily renormalized for comparison to the GCs.
 Peaks are apparent in both tracers at [$Z$/H] $\sim -1.3$ and $-0.6$, providing the first
 demonstration beyond the MW that GC bimodality is reflected in the underlying stellar halo.
 }
   \label{fig:bimod}
\end{center}
\end{figure}

Adding to this picture is recent, deep {\it Hubble Space Telescope} ({\it HST}) imaging in the outer halo of a nearby S0 galaxy,
which detected the elusive second peak of metal-poor red-giant-branch (RGB) stars corresponding to the metal-poor GC peak
(Figure~\ref{fig:bimod}; \cite[Peacock et al.\ 2015]{Peacock15}).
This observation supports the view that blue GC subpopulations trace underlying metal-poor stellar halo populations
and that GC bimodality implies two distinct, fundamental modes of early star formation.
The different relative peak heights require a higher GC formation efficiency or survival rate in the metal-poor population, for
reasons that remain undetermined
(see Harris \& Harris 2002).

A clearer view of GC origins could be obtained through estimates of their ages,
but even in the MW, 
the best current methods are still uncertain at the $\sim$~2~Gyr level --
which is the difference between forming at $z\sim 3$ and $z\sim 20$.
This conundrum motivates an alternative approach where 
GC metallicities $Z$ are used as a proxy for age,
after recognizing that metals build up in galaxies like clockwork
(\cite[Shapiro et al.\ 2010]{Shapiro10}; \cite[Spitler 2010]{Spitler10}).
Thus the GCs can be connected to the 
observed galaxy mass--metallicity--redshift relation
in order to thereby infer both their epochs and their sites of formation.

\begin{figure}[h]
\begin{center}
 \includegraphics[width=4.5in]{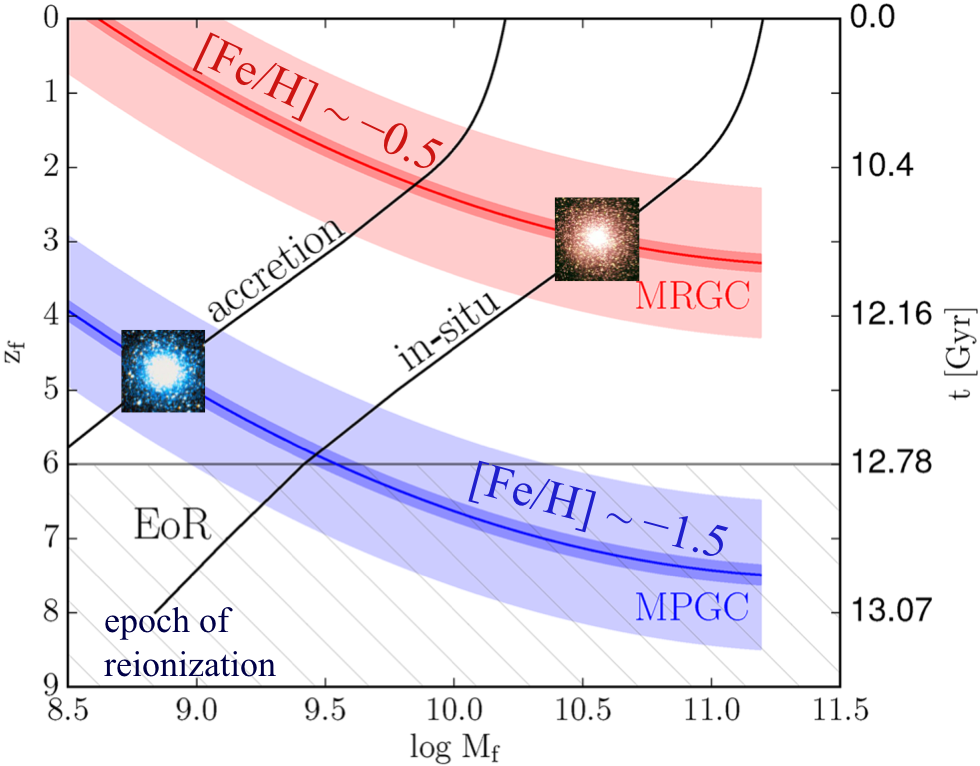} 
 \caption{Formation space of GCs around E/S0 galaxies:
 redshift vs.\ host galaxy stellar mass.
The colored bands show possible solutions for the metal-poor and metal-rich GC subpopulations,
bsaed on their measured metallicities.
The near-orthogonal curves show example growth tracks, both for the main galaxy and its satellites.
Metal-rich GCs are inferred to form at typically $z \sim 3$ in high-mass galaxies,
and metal-poor GCs at $z \sim 5$ in dwarfs.
}
   \label{fig:metal}
\end{center}
\end{figure}

We have applied this ``metallicity matching'' method to GC spectroscopic data for a sample of E/S0s from SLUGGS
(\cite[Forbes et al.\ 2015]{Forbes15}).
As  Figure~\ref{fig:metal} shows, for each GC subpopulation, there is a range of possible average formation epochs:
for example, the metal-poor GCs could have been formed very early in massive galaxies, or later in dwarf galaxies.
This degeneracy is broken by invoking plausible growth curves for the GC host galaxies:
one curve for the main progenitor galaxy, and one for a typical 1:10 mass-ratio accreted satellite.
Assuming that the metal-rich GCs are a predominantly in situ population, they appear to have
formed in the range $z \sim$~2--4.
If the metal-poor GCs are primarily accreted, then they formed at $z \sim$~4--5.

This relatively recent birth date for the metal-poor GCs would make them a post-reionization population --
an important conclusion since it has been proposed that reionization either produced
a hiatus between the two GC subpopulations, or was caused by the GCs themselves
(e.g., \cite[Beasley et al.\ 2002]{Beasley02}; \cite[Katz \& Ricotti 2013]{Katz13}).
On the other hand, there are significant caveats.  One is that the galactic metallicity relation is
not well constrained at redshifts above $z \sim 4$.
Another is that there may have also been an earlier, in situ phase of metal-poor GC formation in the main galaxy,
which is reflected in some problems of crafting a self-consistent pure-accretion model
(e.g., \cite[Spitler et al.\ 2012]{Spitler12})).

The inferred metal-rich GC birth date is more secure, and raises intriguing possibilities for further progress since
the Universe at $z\sim3$ is becoming well studied, as the ``high noon'' of cosmic star formation.
We can now see directly where and how the intense bursts of star formation activity occurred that must
have formed the bulk of the metal-rich GC populations.
The answer is remarkable:
rather than the gas-rich major mergers that dominated the thinking about GC and galaxy formation for many years,
high-$z$ star formation is seen to concentrate in extended, turbulent, gas-rich galactic disks
(e.g., \cite[Genzel et al.\ 2008]{Genzel08}). 
These disks host supergiant clumps of gas and young stars that seem natural spawning sites for GCs
(Shapiro et al.\ 2012; Kruijssen 2015).

This consolidation of observations at low and high redshifts motivates a new generation of theoretical models
for the assembly of GC systems in a cosmological context.
The models should include violent disk instabilities (e.g., \cite[Porter et al.\ 2014]{Porter14}), and
should respect the full range of chemo-dynamical constraints:
not only the numbers, ages, and metallicities  of GCs (\cite[Tonini 2013]{Tonini13}; \cite[Li \& Gnedin 2014]{Li14}), 
but also their spatial and orbital distributions -- which have so far posed severe theoretical challenges
(see Brodie et al.\ 2014, Section 5.4, for more discussion).

\section{Compact stellar systems: don't mind the gaps}\label{sec:CSS}

UCDs were discovered serendipitously through spectroscopic surveys
(\cite[Hilker et al.\ 1999]{Hilker99}; \cite[Drinkwater et al.\ 2000]{Drinkwater00}).
They appear similar to GCs but typically have larger sizes (half-light radii $r_{\rm h} \sim$~10--40~pc
rather than $\sim$~2--4~pc) and higher luminosities (up to $\sim 10^7 L_\odot$).
Their formation mechanisms may 
include merging super star clusters 
and tidal stripping of galactic nuclei.
However, their distributions in basic parameter space and in galactic environments are not clear.
Two major efforts to map out the properties of UCDs more systematically
include SLUGGS (\S\ref{sec:cosmo})
and the Archive of Intermediate Mass Stellar Systems project (AIMSS; \cite[Norris et al.\ 2014]{Norris14}).

\begin{figure}[h]
\begin{center}
 \includegraphics[width=\textwidth]{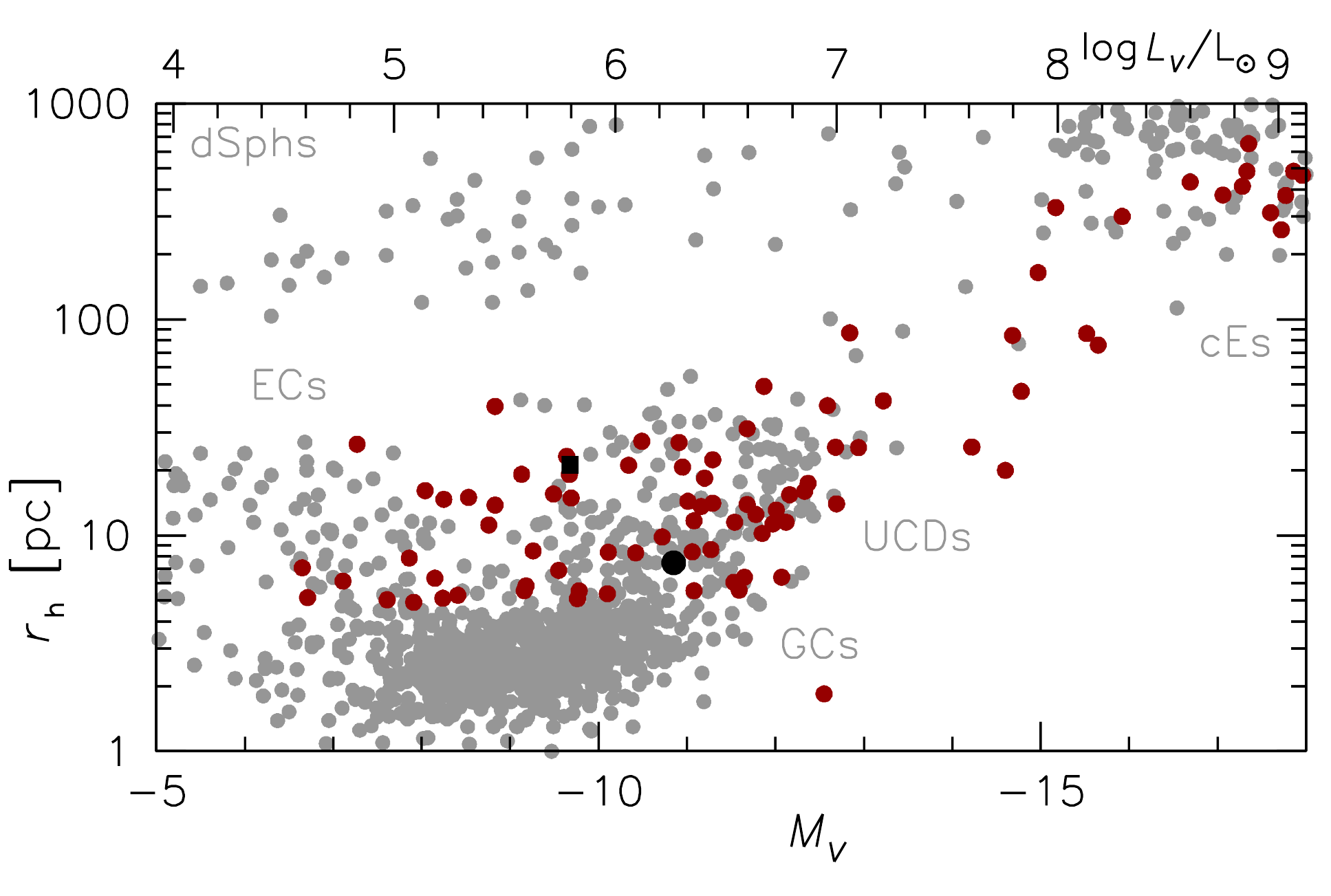} 
 \caption{Half-light radius vs.\ $V$-band absolute magnitude for distance-confirmed hot stellar sytems.
 Larger red circles mark objects discovered in the past 2 years.
 The black square and circle show NGC~2419 and $\omega$~Cen, respectively.
}
   \label{fig:size}
\end{center}
\end{figure}

These surveys revealed that UCDs are {\it not} synonymous with being very luminous: the 
family of large-$r_{\rm h}$ objects also extends to lower luminosities and meets up with the ``faint fuzzies''
or extended clusters
(\cite[Brodie et al.\ 2011]{Brodie11}; \cite[Forbes et al.\ 2013]{Forbes13})
This expansion of classification boundaries means that the unusual MW object NGC~2419 is the nearest UCD.
At the other end of the spectrum, record-breaking dense stellar systems have been found
(\cite[Strader et al.\ 2013]{Strader13}; \cite[Sandoval et al.\ 2015]{Sandoval15}), 
with these novel extremes of high- and low-surface brightness 
demonstrating the pivotal role of selection effects in shaping our conceptions of stellar systems.
An additional uncharted area of parameter space is the gap between UCDs and compact ellipticals, which
is now seen to be filled in with cE--UCD transition objects
(see size--luminosity diagram in Figure~\ref{fig:size}\footnote{Compilation maintained at {\tt http://sages.ucolick.org/spectral\_database.html}}).

This steadily growing menagerie of diverse compact stellar systems compounds the need to decipher their interrelations and origins.
The UCDs probably include more than one underlying subpopulation, and recent cosmological models conclude
that most of them are unusually large star clusters, with only a minority being stripped nuclei
(\cite[Pfeffer et al.\ 2014]{Pfeffer14}).
Tests of these claims are needed from in-depth observational studies of the UCDs. 
Some cases of stripped nuclei are obvious, based on an overmassive central black hole,
or on an extended star formation history
(\cite[Seth et al.\ 2014]{Seth14}; \cite[Norris et al.\ 2015]{Norris15}).
Metallicity is another useful diagnostic, as cEs and the highest-mass UCDs were found to be relatively metal-rich,
in a probable reflection of stripped nuclei origins
(\cite[Janz et al.\ 2016]{Janz16}).
Further hereditary clues may come from kinematics and dynamics
(\cite[Brodie et al.\ 2011]{Brodie11}; \cite[Strader et al.\ 2011]{Strader11}; \cite[Zhang et al.\ 2015]{Zhang15}), and from
more detailed abundances,
such as the very high nitrogen and sodium levels seen in some massive UCDs 
(\cite[Strader et al.\ 2013]{Strader13}; \cite[Sandoval et al.\ 2015]{Sandoval15}).

\section{Ultra-compact dwarfs in formation}\label{sec:inform}

\begin{figure}[ht!]
\begin{center}
\includegraphics[width=\textwidth]{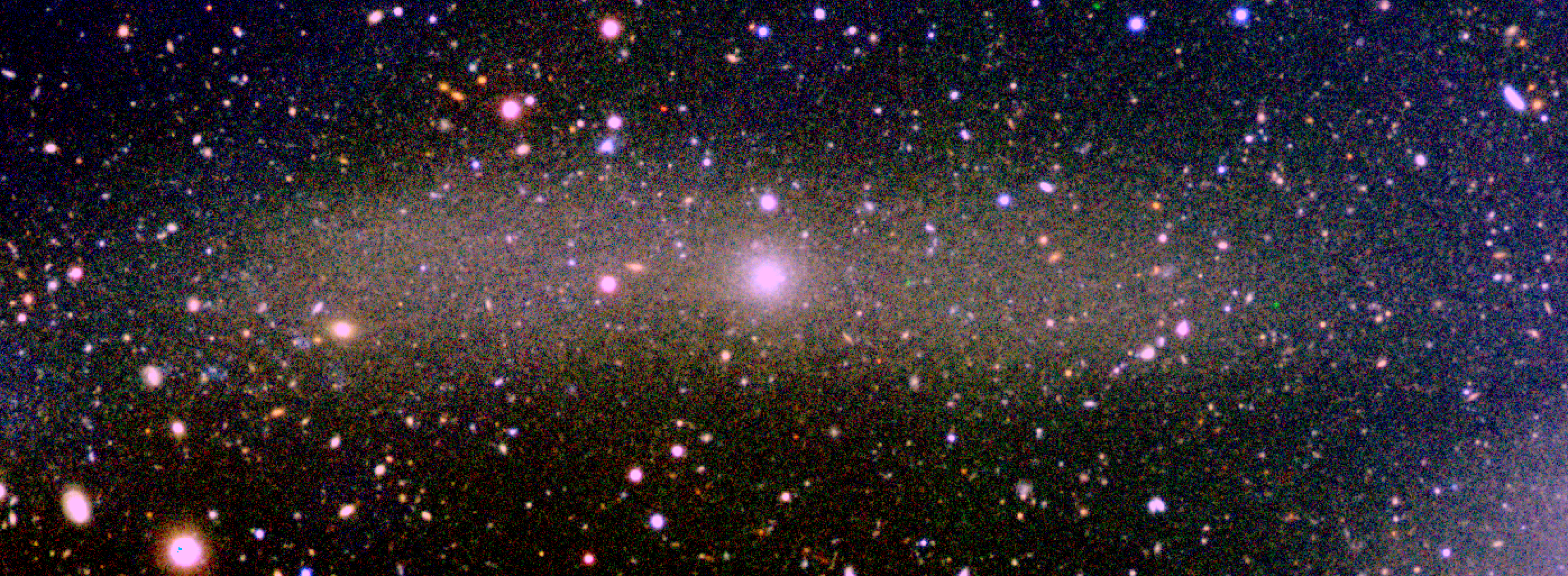} 
 \caption{Image of the ``plume'' section of the tidal stream around NGC~3628,
 from Subaru/Suprime-Cam ($20\times7$~kpc image section, $gri$ color composite).
 A central compact stellar cluster is visible, with properties similar to $\omega$~Cen.
}
   \label{fig:n3628}
\end{center}
\end{figure}

Besides the clues to the past of UCDs, it is also possible to witness ongoing UCD formation by tidal stripping.
The nearby ``Hamburger'' spiral galaxy NGC~3628 boasts a tidal stream that
appears to contain the nucleus of the disrupting dwarf galaxy
(Figure~\ref{fig:n3628}; \cite[Jennings et al.\ 2015]{Jennings15}).
This object has a luminosity of $\sim 10^6 L_\odot$, a size of $\sim 10$~pc, and a metallicity of
[$Z$/H]~$\sim -0.8$~dex
(from Keck/ESI and Large Binocular Telescope / MODS spectroscopy).
It seems destined to become a massive stellar system like $\omega$~Cen.  
Other, similar examples have also been found recently
(e.g., \cite[Foster et al.\ 2014]{Foster14}), and raise the possibility of providing 
empirical information about the rates and dynamics
of GC/UCD formation by tidal stripping.

\acknowledgements

Supported by National Science Foundation awards AST-1109878,  AST-1211995, AST-1515084, and AST-1518294.


\begin{thebibliography}{}

\bibitem[Beasley et al.(2002)]{Beasley02} Beasley, M.~A., Baugh, C.~M., Forbes, D.~A., Sharples, R.~M., 
\& Frenk, C.~S.\ 2002, {\it MNRAS}, 333, 383 

\bibitem[Brodie et al.(2011)]{Brodie11} Brodie, J.~P.,  Romanowsky, A.~J., Strader, J., \& Forbes, D.~A.\ 2011, {\it AJ}, 142, 199 

\bibitem[Brodie et al.(2012)]{Brodie12} Brodie, J.~P., Usher,  C., Conroy, C., et al.\ 2012, {\it ApJ} (Letters), 759, L33 

\bibitem[Brodie et al.(2014)]{Brodie14} Brodie, J.~P.,  Romanowsky, A.~J., Strader, J., et al.\ 2014, {\it ApJ}, 796, 52

\bibitem[Drinkwater et al.(2000)]{Drinkwater00} Drinkwater, M.~J.,  Jones, J.~B., Gregg, M.~D., \& Phillipps, S.\ 2000, {\it PASA}, 17, 227 

\bibitem[Forbes et al.(2013)]{Forbes13} Forbes, D.~A., Pota, V.,  Usher, C., et al.\ 2013, {\it MNRAS}, 435, L6

\bibitem[Forbes et al.(2015)]{Forbes15} Forbes, D.~A., Pastorello, N., Romanowsky, A.~J., et al.\ 2015, {\it MNRAS}, 452, 1045 

\bibitem[Foster et al.(2014)]{Foster14} Foster, C., Lux, H., Romanowsky, A.~J., et al.\ 2014, {\it MNRAS}, 442, 3544 

\bibitem[Genzel et al.(2008)]{Genzel08} Genzel, R., Burkert, A., Bouch{\'e}, N., et al.\ 2008, {\it ApJ}, 687, 59 

\bibitem[Harris  \& Harris(2002)]{Harris02} Harris, W.~E., \& Harris, G.~L.~H.\ 2002, {\it AJ}, 123, 3108 

\bibitem[Hilker et al.(1999)]{Hilker99} Hilker, M., Infante, L., Vieira, G., Kissler-Patig, M., \& Richtler, T.\ 1999, {\it A\&AS}, 134, 75 

\bibitem[Janz et al.(2016)]{Janz16} Janz, J., Norris, M.~A., Forbes, D.~A., et al.\ 2016, {\it MNRAS}, in press, arXiv:1511.03264 

\bibitem[Jennings et al.(2015)]{Jennings15} Jennings, Z.~G., 
Romanowsky, A.~J., Brodie, J.~P., et al.\ 2015, {\it ApJ} (Letters), 812, L10 

\bibitem[Katz \& Ricotti(2013)]{Katz13} Katz, H., \& Ricotti, M.\ 2013, {\it MNRAS}, 432, 3250 

\bibitem[Kruijssen(2015)]{Kruijssen15} Kruijssen, J.~M.~D.\ 2015, {\it MNRAS}, 454, 1658 

\bibitem[Larsen et al.(2001)]{Larsen01} Larsen, S.~S., Brodie,  J.~P., Huchra, J.~P., Forbes, D.~A., 
\& Grillmair, C.~J.\ 2001, {\it ApJ}, 121, 2974 

\bibitem[Li \& Gnedin(2014)]{Li14} Li, H., \& Gnedin, O.~Y.\ 2014, {\it ApJ}, 796, 10 

\bibitem[Norris et al.(2014)]{Norris14} Norris, M.~A., Kannappan, S.~J., Forbes, D.~A., et al.\ 2014, {\it MNRAS}, 443, 1151

\bibitem[Norris et al.(2015)]{Norris15} Norris, M.~A., Escudero, C.~G., Faifer, F.~R., et al.\ 2015, {\it MNRAS}, 451, 3615

\bibitem[Peacock et al.(2015)]{Peacock15} Peacock, M.~B.,  Strader, J., Romanowsky, A.~J., \& Brodie, J.~P.\ 2015, {\it ApJ}, 800, 13 

\bibitem[Pfeffer et al.(2014)]{Pfeffer14} Pfeffer, J., Griffen,  B.~F., Baumgardt, H., \& Hilker, M.\ 2014, {\it MNRAS}, 444, 3670 

\bibitem[Porter et al.(2014)]{Porter14} Porter, L.~A., Somerville, R.~S., Primack, J.~R., \& Johansson, P.~H.\ 2014, {\it MNRAS}, 444, 942 

\bibitem[Pota et al.(2013)]{Pota13} Pota, V., Forbes, D.~A., Romanowsky, A.~J., et al.\ 2013, {\it MNRAS}, 428, 389 

\bibitem[Sandoval et al.(2015)]{Sandoval15} Sandoval, M.~A., Vo, R.~P., Romanowsky, A.~J., et al.\ 2015, {\it ApJ} (Letters), 808, L32 

\bibitem[Searle \& Zinn(1978)]{Searle78} Searle, L., \& Zinn, R.\ 1978, {\it ApJ}, 225, 357 

\bibitem[Seth et al.(2014)]{Seth14} Seth, A.~C., van den Bosch, R., Mieske, S., et al.\ 2014, {\it Nature}, 513, 398

\bibitem[Shapiro et al.(2010)]{Shapiro10} Shapiro, K.~L., Genzel,  R., F\"orster Schreiber, N.~M.\ 2010, {\it MNRAS}, 403, L36 

\bibitem[Spitler(2010)]{Spitler10} Spitler, L.~R.\ 2010, {\it MNRAS}, 406, 1125 

\bibitem[Spitler et al.(2012)]{Spitler12} Spitler, L.~R.,  Romanowsky, A.~J., Diemand, J., et al.\ 2012, {\it MNRAS}, 423, 2177

\bibitem[Strader et al.(2011)]{Strader11} Strader, J., Romanowsky, A.~J., Brodie, J.~P., et al.\ 2011, {\it ApJS}, 197, 33 

\bibitem[Strader et al.(2013)]{Strader13} Strader, J., Seth, A.~C., Forbes, D.~A., et al.\ 2013, {\it ApJ} (Letters), 775, L6

\bibitem[Tonini(2013)]{Tonini13} Tonini, C.\ 2013, {\it ApJ}, 762, 39 

\bibitem[Usher et al.(2012)]{Usher12} Usher, C., Forbes, D.~A.,  Brodie, J.~P., et al.\ 2012, {\it MNRAS}, 426, 1475 

\bibitem[Zepf \& Ashman(1993)]{1993MNRAS.264..611Z} Zepf, S.~E., \& Ashman, K.~M.\ 1993, {\it MNRAS}, 264, 611

\bibitem[Zhang et al.(2015)]{Zhang15} Zhang, H.-X., Peng, E.~W., C{\^o}t{\'e}, P., et al.\ 2015, {\it ApJ}, 802, 30

\end{thebibliography}
\end{document}